\documentclass{article}
\usepackage{spconf,amsmath,graphicx}
\usepackage{multirow}
\def\x{{\mathbf x}}
\def\L{{\cal L}}

\title{Analyzing Large Receptive Field Convolutional Networks for\\Distant Speech Recognition}
\name{Salar Jafarlou, Soheil Khorram, Vinay Kothapally, John H.L. Hansen}
\address{Center for Robust Speech Systems (CRSS), Erik Jonsson School of Engineering \& Computer Sciences\\ The University of Texas at Dallas, Richardson,TX, USA}
\begin{document}
%
\maketitle
\begin{abstract}
 Despite significant efforts over the last few years to build a robust automatic speech recognition (ASR) system for different acoustic settings, the performance of the current state-of-the-art technologies significantly degrades in noisy reverberant environments.
 Convolutional Neural Networks (CNNs) have been successfully used to achieve substantial improvements in many speech processing applications including distant speech recognition (DSR). However, standard CNN architectures were not efficient in capturing long-term speech dynamics, which are essential in the design of a robust DSR system. In the present study, we address this issue by investigating variants of large receptive field CNNs (LRF-CNNs) which include \emph{deeply recursive networks}, \emph{dilated convolutional neural networks}, and \emph{stacked hourglass networks}. To compare the efficacy of the aforementioned architectures with the standard CNN for Wall Street Journal (WSJ) corpus, we use a hybrid DNN-HMM based speech recognition system. We extend the study to evaluate the system performances for distant speech simulated using realistic room impulse responses (RIRs). Our experiments show that with fixed number of parameters across all architectures, the large receptive field networks show consistent improvements over the standard CNNs for distant speech. Amongst the explored LRF-CNNs, stacked hourglass network has shown improvements with a 8.9\% relative reduction in word error rate (WER) and 10.7\% relative improvement in frame accuracy compared to the standard CNNs for distant simulated speech signals.

 
\end{abstract}
\section{Introduction}
\label{sec:intro}

Distant Speech Recognition (DSR) is a technology that uses distant microphone(s) to accomplish natural human-machine interfaces. Recent years have seen the application of DSR in consumer devices, such as Amazon Echo, Google Home, smart TVs, etc. Due to the existence of background noise, multiple overlapping speakers and reverberation, building a robust DSR system has become a challenging task for present speech systems. Broadly speaking, a DSR system can be split into two sub-tasks: (i) a front-end speech enhancement system, and (ii) a back-end automatic speech recognition (ASR) system which can be designed to operate on speech recordings from either a single distant microphone or multiple distant microphones. A DSR system, engineered using multiple distant microphones, use advanced front-end microphone array processing techniques that yield in a substantially reduced word error rate (WER) compared to systems engineered using a single distant microphone. Most back-end state-of-the-art ASR systems used in a DSR system typically divide the recognition task into three sub-tasks: (i) feature extraction, (ii) acoustic modeling, and (iii) language modeling, which are optimized independently to achieve the best performance. 


Over the years, steady attempts by speech community researchers have helped in optimizing the aforementioned building blocks of the ASR system. Feature extraction, a process of extracting discriminative characteristics from speech signals to accurately classify linguistic content has been extensively studied, leading in features such as Mel-filterbank cepstral coefficients (MFCCs) and perceptual linear prediction coefficients (PLPs) providing optimum efficiency for many speech-related systems. Similarly, extensive studies in natural language processing (NLP) have shown that recurrent neural network-based language models (RNN-LMs) generate accurate probability distributions over word sequences, helping an ASR system to decrease prediction errors. For acoustic modeling, researchers have used Hidden Markov Models (HMMs) and Gaussian Mixture Models (GMMs) for more than a decade. Later, studies in this area have shown that acoustic models based on fully connected deep neural network (FC-DNNs) outperformed the conventional GMM-HMM systems. In addition, significant improvements were also made by replacing fully connected DNNs with convolutional neural networks (CNNs) because of their effectiveness in capturing local (short-term) dependencies of speech signals. This leads to significant improvements in WER for speech recordings from a close-talk microphone. Consequently, CNNs do not efficiently capture global (long-term) dependencies which make them less effective in designing a DSR system.

CNN is a multi-layer stacked neural network which includes convolutional layers, non-linearities, and pooling layers(in some frameworks)~\cite{krizhevsky2012imagenet}. Convolutions in different layers of the standard CNN consider current and few neighboring inputs from a previous layer to produce a single output. As the number of layers in this network increases the region of the input space (the first layer of the network) that affects a neuron in a particular layer of the neural network also increases. This region is well recognized in CNN architecture as the receptive field. In general, any neuron of any layer can be investigated for its receptive field. Nonetheless, this term is commonly used to describe the region of an input that impacts a specific network output. Therefore, we can say that the receptive field of a CNN is a measure of its temporary learning capacity that increases linearly with the number of layers and the size of a convolution kernel used in a CNN. In CNNs, it is evident that the receptive field size can be increased in the following ways: (i) stacking more layers (increasing the depth of the network), (ii) sub-sampling (introducing pooling after convolutions, having a lower stride), and (iii) increasing kernel size (dilating the convolutional kernel). Although the expansion of the receptive field significantly increases the number of parameters, it is beneficial in capturing global and local dependencies which are crucial for building a DSR system.

The goal of this paper is to explore the efficiency of DSR systems built using hybrid DNN-HMM and large receptive field networks for acoustic models. We perform a thorough analysis on the design of these networks and on the relationship between receptive field size and the number of parameters of the networks.
\vspace{-1em}

\section{Related Work}
\label{sec:prevwork}

In this section, we discuss the past and present research work relevant to capturing long-term dependencies in speech. There are two approaches to address the concern of capturing long-term dependencies in a speech signal described in the previous section: (i) using feature extraction techniques that take into account long-term dependencies while extracting features, or (ii) using acoustic models that can learn the long-term dependencies given short-term speech features \cite{Poveytdnn}.

Initial efforts from researchers in speech and audio processing were exclusive to explore feature extraction strategies to address this issue. For instance, (i) TRAPs, a feature extraction technique which replaced standard spectral patterns with long-term temporal patterns of spectral energies \cite{hermansky1999temporal}, (ii) A wavelet-based multi-scale spectro-temporal feature extraction technique which consider multiple time and spectral resolutions tuned to capture fast and slow changes in modulation patterns \cite{mesgarani2004speech}, and more recently (iii) Features from deep scattering spectrum, which extend standard MFCCs by calculating multiple-orders of modulation spectrum coefficients with the use of wavelet cascades \cite{anden2014deep,yousefi2019probabilistic}. These long-term speech dynamics capturing features showed reasonable performance improvements when tailored to a specific task (or) speech from a particular acoustic environment. These feature extraction techniques can not be generalized for all acoustic conditions because it needs the expertise to tune parameters in the extraction process to compensate for the distortions induced by an acoustic condition on speech which are inconsistent and change swiftly. It was therefore found that the best approach to address long-term speech dynamics capturing problem may be to seek for alternative strategies for acoustic modeling rather than the feature extraction. Later, acoustic modeling strategies were researched in great detail to deal with this problem.

With advances in machine learning, FC-DNNs learning strategies were adapted to build robust state-of-the-art acoustic models that can statistically map an acoustic sound precisely to its corresponding transcript. Although FC-DNNs have shown significant improvements over GMM-HMM-based acoustic modeling, their temporal modeling capabilities were limited as they operate on the information from a fixed-size sliding window of acoustic frames. This made them unsuitable for handling long-term dependencies. Subsequently, recurrent neural networks (RNNs), a progression to FC-DNNs with cyclic connections over time, were able to collect and store information for an arbitrary number of neighboring acoustic frames, showing their capacity to capture long-term dependencies \cite{greff2017lstm}. Several RNN architectures have since been explored for acoustic modeling (e.g., GRUs~\cite{wu2016investigating}, LSTMs~\cite{graves2013speech, ghorbani2018advancing}, BLSTMs~\cite{graves2013hybrid}, RNMs \cite{RNM}). Training RNNs are usually performed through a time-expansion operation where the input at time `$t+1$' relies on the output at time `$t$'. Due to this time-expansion operation, parallelization of training routines for these networks becomes quite challenging even with techniques such as sequence batching and distributed optimization.

Convolutional neural networks are one of many other machine learning strategies adapted for acoustic modeling to handle the long-term dependencies in ASR. Similar to RNNs, CNNs have also shown significant improvements in ASR performance over FC-DNNs \cite{CNNforASR, ExploringCNN,ghorbani2019Domain,ghorbani2018leveraging}. Recent research also shows that the use of residual connections can train deeper CNN architectures in a more efficient way compared to RNNs \cite{ResidualImageRecognition}. Thus, deep CNNs with restricted local connectivity and weight sharing were successfully used in document recognition \cite{document}. Researchers have studied various variants of CNNs that use the concept of large receptive field to build robust systems in the areas of human pose recognition \cite{newell2016stacked}, face expression recognition \cite{yang2017stacked}, human speech emotion recognition \cite{khorram2017capturing, khorram2019jointly}, signature verification \cite{al2014network} and also in many machine learning applications associated with super-resolution image processing. Tiled-CNNs that learns rotational and scale-invariant features over time has proven to perform better than traditional CNN for small time-series data \cite{TiledCNNs}. CNNs have also been used for speech dereverberation applications in multiple configurations and have successfully demonstrated their ability to learn the long-term effects of reverberation on speech \cite{speechDereverb, yousefi2018assessing}. In addition, Dilated CNNs have also proven their abilities to learn relevant information from a bigger context \cite{dilatedCNNs}. Therefore, we focus on studying the large-receptive field networks for acoustic modeling, especially for distant speech recognition. 
\vspace{-1em}

\section{Methods}
\label{sec:methods}

In this section, we discuss the working principles of standard CNN, dilated CNNs, and stacked hourglass network. We compute and compare the receptive field size of the mentioned networks to better understand the increase/decrease in the performance of each network.

\begin{figure*}[t!]
  \centering
  \includegraphics[height=13cm,width=\linewidth]{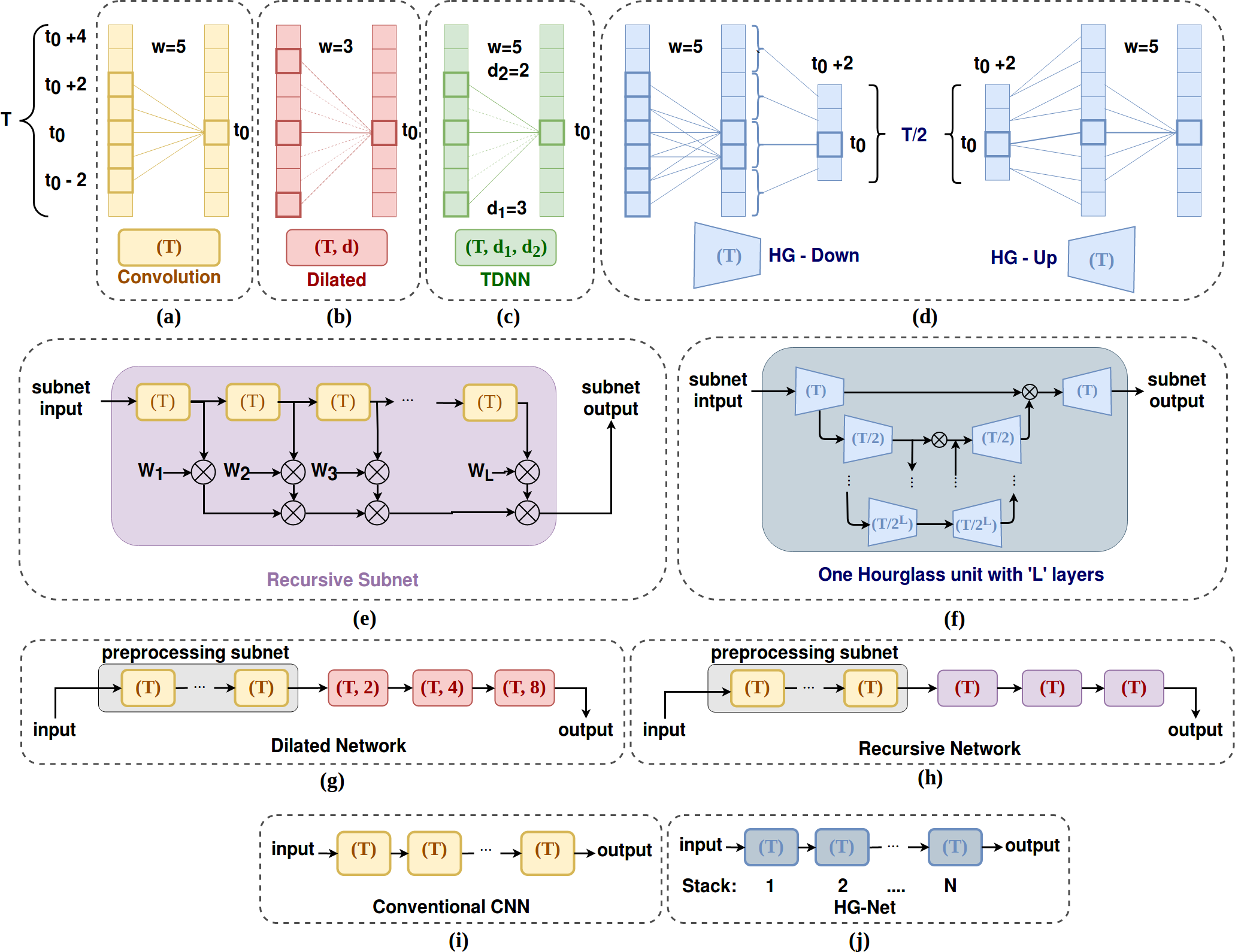}
  \caption{(a-d) Various convolutional filters used in convolutional neural networks, (e-f) Illustrates a single unit of recursive and hourglass networks that are stacked to build respective networks. (g-i) Fully stacked network architectures.\vspace{-0.7cm}}
  \label{fig:layersdiagram}
  \vspace{1em}
\end{figure*}

\subsection{Standard CNN}
As mentioned in the previous sections, CNNs can be considered as a variation of regular feed-forward networks. In CNNs, \textit{weight sharing} is normally achieved by sliding a linear filter throughout the output of the previous layer, see Fig-\ref{fig:layersdiagram}(a). A conventional CNN is built by stacking up `$L$' convolutional layers. Assuming each convolutional layer uses a linear filter of kernal width `$W$', we can compute the receptive field as follows: 
\begin{equation}
RF_{standard} = L (W - 1)+1
\end{equation}

where $RF_{CNN}$ is the RF size of the standard CNN. It is evident from this equation that the RF size increases linearly with respect to both $W$ and $L$. As the RF size increases, the number of learning parameters also increases linearly, making the network not effective for tasks where large receptive field sizes are required. Also, due to the linear relationship between RF size and network complexity, it is difficult to find a trade-off point.

\subsection{Dilated Networks (DIL-Net) }


Networks that use dilated convolutions have shown to be effective in many tasks, including image segmentation~\cite{yu2015multi}, speech synthesis~\cite{oord2016wavenet} and ASR~\cite{sercu2016dense}. Dilated networks, provide an effective technique for increasing the RF size without causing a significant rise in the number of learning parameters. In a dilated network, the convolutional filter (kernel) is obtained by inserting zeros between the regular filter samples. This method expands the filter in time at the expense of lower resolution; making the filter sparse when compared to a standard CNN convolutional filter. Fig-\ref{fig:layersdiagram}(b) shows an example of the dilated convolution filter with the dilation factor of $d=3$. A dilated convolutional filter is simply obtained by inserting $(d-1)$ zeros symmetrically between successive filter coefficients. A dilated network is generally constructed by stacking $n$ dilated convolutional layers with a $2^n$ dilation factor for each layer and a preprocessing subnet at the beginning. A preprocessing subnet is a feature processing block built using a stack of regular convolutional layers, see Fig-\ref{fig:layersdiagram}(g). The RF size of this dilated network can be computed as follows:

\begin{equation}
RF_{dilated}=(L+(2^{L-1}-1))(W-1)+1
\end{equation}

where $L$ is the number of layers and $W$ is the width of the convolutional layers. 
The RF size grows exponentially with the number of layers, while the number of parameters grows linearly.

A variant of dilated networks is achieved by inserting zeros asymmetrically between 
successive filter coefficients ~\cite{peddinti2015time}. This network is commonly known as time-delay neural network (TDNN). Fig-\ref{fig:layersdiagram}(c) shows a single layer of TDNN with asymmetric dilations. Each layer in a TDNN can have different dilation values $d_{l,1}$ and $d_{l,2}$. The asymmetric dilation characteristic of TDNN makes it more flexible and gives the network a better learning capacity compared to dilated networks. On the contrary, $(d_{l,1},d_{l,2})$ hyper-parameters are extremely data-dependent and can only be tuned by empirical studies to optimize the efficiency of the networks. The RF size of a TDNN can be computed as follows:

\begin{equation}
RF_{tdnn}=1 + \sum_{l=1}^{L}{(d_{l,1}+d_{l,2})}
\end{equation}

\subsection{Stacked Hourglass Network (HG-Net)}
Stacked hourglass structure (HG-Net) was initially designed to solve facial landmark localization \cite{yang2017stacked} and human pose estimation \cite{newell2016stacked} which need to process both high-resolution (local view) and low-resolution (global view) versions of an image in parallel~\cite{oliva2006building}. This property is equivalent to processing short-term and long-term temporal dynamics of the speech signal. 

HG-Net is build using a stack of hourglass networks to processes both short-term and long-term temporal dependencies in parallel, see Fig-\ref{fig:layersdiagram}(j). As shown in Fig-\ref{fig:layersdiagram}(f), each hourglass unit in an HG-net contains `$L$'-layers with two sub-networks: (1) a down-sampling network; (2) an up-sampling network in each layer. Fig-\ref{fig:layersdiagram}(d) shows the convolutions involved in the down/up sampling networks. The down-sampling network generates low-resolution representations of the input, and the up-sampling network converts the representations learned from low-resolution to high-resolution signals. 

The down-sampling network consists of a series of convolutions and max-pooling layers. The max-pooling layer reduces the resolution of the signal and increases the RF of the network. Various pooling operations can be used instead of max-pooling. The up-sampling network consists of a series of up-pooling and convolutional layers. This network combines all the representations learned from different resolutions of input. In addition, the hourglass network exploits a specific skip connection mechanism that connects representations than can allow us to leverage many layers for down-sampling and up-sampling networks without having the vanishing gradient problem. Therefore, we can down-sample the input signal to a low resolution to achieve a large RF.

Assuming $W_{d}, P_{d}, L_{d}$ to be the filter size of convolutions, pooling and number of layers in a down-sampling network\footnote{The number of layers in the down-sampling and up-sampling networks must be equal in the hourglass network}, the RF size of a down-sampling network can be computed as:
\begin{equation}
RF_{down} = L_{d}(W_{d}+P_{d}-1)-1
\end{equation}

RF size of the stacked hourglass network (HG-Net), $RF_{stacked-hg}$, can be approximately calculated as:
\begin{equation}
RF_{stacked-hg} \approx S\times(RF_{down}*2^{L})\vspace{-0.2cm}
\end{equation}
where $S,L$ denotes the number of hourglass units in an HG-Net and number of layers in each hourglass unit. This shows that $RF_{stacked-hg}$ exponentially increases with $L$. RF size can be efficiently increased by using more layers in the down-sampling and up-sampling networks as well.

\subsection{Deeply Recursive Network (REC-Net)}
Deeply recursive neural network (REC-Net) is first proposed by Kim et al. as an image super-resolution method~\cite{kim2016deeply}. The idea is to use a big network with a large number of layers and allow different layers to share their learnable parameters. REC-Net is a stack of recursive subnetworks, as it is shown in Figure 1(h). Each recursive network (Figure 1(e)) contains a series of convolutional layers (Figure 1(a)) that all of them share the same weights. In the recursive subnetwork, increasing the number of layers will increase the RF size without increasing the number of parameters. REC-Net can provide a large RF with a small number of parameters. 

REC-Net has a number of problems: (1) to capture a large RF, we must use a large stack of identical layers in the recursive subnetwork. Training this structure is difficult and may lead to a vanishing/exploding gradient problem. To solve this problem, authors in ~\cite{kim2016deeply} proposed a skip-connection strategy shown in Figure 1(e) where the output of the recursive subnetwork is obtained through a weighted average of the output of all layers in the recursive subnetwork; (2) training REC-Net is computationally expensive (in both time and memory requirements) since this network requires a large number of identical layers to capture long-term dependencies.

Unlike the conventional network, all these big receptive field networks provide an efficient way to increase the size of the receptive field without causing a significant rise in the number of learning parameters. Thus, we compare the efficiency  of these networks with the conventional network by setting the number of learning parameters to be the same across all networks.
\vspace{-1em}

\section{Experimental Setup}
\subsection{Distant Speech Simulation}
Reverberation in distant speech recordings can be simulated by convolution of the audio signals with a room impulse response from a point source to a receiver location in a room. RIRs are highly sensitive to changes in receiver position, speaker position or positions of different obstacles in the room \cite{RoomAcoustics}. Assuming the RIRs do not change over a small instances of time corresponding to a particular source and receiver positions, We use a set of 325 real RIRs composed of three databases: the RWCP sound scene database \cite{RWCP_RIR}, the REVERB challenge database \cite{REVERB_RIR} and the Aachen impulse response database \cite{AIR_RIR} and clean speech signals from WSJ corpus to simulate the distant speech recordings.

\subsection{Training LRF Networks}
We used Wall Street Journal (WSJ) dataset to evaluate the performance of the large RF convolutional networks explained in the previous section. The training data consists of 80 hours of speech both telephone and microphone speech, the bulk of which is in English. All wideband audio is downsampled to 8kHz. The evaluation is performed on the Eval93 subset of the WSJ. The Dev93 subset of the WSJ is used to tune the parameters across all networks. We used 40-dimensional Mel-filterbank (MFB) features normalized with Cepstral Mean and Variance Normalization (CMVN) as the input features of the networks~\cite{khorram2018priori, zhang2019exploiting}. We also implemented Feature space Maximum Likelihood Linear Regression (FMLLR) transformation in our initial experiments, but it did not yield performance improvements. Since the main focus of this paper is on the effect of large RF covering, we did not explore the effect of speaker normalization methods (e.g., i-vectors) in our experiments. We trained our models up to 20 epochs using the Adam optimizer ($\alpha = 0.001$). Our initial experiments showed that ReLu activation function outperforms other activations and therefore we applied ReLu in the intermediate layers of the networks. We employed softmax for the output layer. We also implemented a discriminative softmax (AMSoftmax)~\cite{Wang_2018_amsoftmax} that did not improve the results.
We trained a triphone model with 3392 states in four iterations and used it as the HMM component of the DNN-HMM pipeline ASR. No language model refinement was applied in the decoding phase. We used Kaldi~\cite{povey2011kaldi} implementation of HMM and we implemented all the networks using the TensorFlow~\cite{abadi2016tensorflow} open-source library. We performed hyper-parameter tuning by leveraging two well-known measures: frame accuracy (Acc) and cross-entropy (CE). In addition to these measures, we also report word error rate (WER) of all networks.

For the standard CNN, we evaluated all the networks with the kernel size of $w = 3$ and $5$, and the number of layers ranging from $L=3$ to $10$. $W=5$ and $L =10$ performed the best in both validation accuracy and WER. As we used raw MFB features, we considered a stack of standard convolutional layers (with 3 layers) as the \textit{preprocessing sub-network} in DIL-Net and REC-Net (Figure 1(g), (h)). We implemented DIL-Net as shown in Figure 1(e). Our DIL-Net contained $3$ and $4$ dilation layers, with the dilation factor ranging exponentially from $2$ to $8$ (i.e., $d=(2,4,8)$). Skip connections were applied to this structure, but they did not lead to better performance. For REC-Net, we used $5$ layers of inner convolutions and $5$ recursive sub-networks. For HG-Net, we validated for the number of stacks $S=1$ to $5$, convolutional kernel size $W = 3$ and $5$ and number of layers $L = 3$ and $5$. Parameters of $S=5, W=5$, and $L=3$ achieved the best performance in terms of WER. For consistency of comparisons, we used the same number of kernels (512 kernels) for all the convolutional layers.\vspace{-0.2cm}

\section{Results }
 In this section, we begin by studying perceptual and objective speech quality measures such as signal-to-noise (SNR), perceptual evaluation speech quality (PESQ), Itakura-Saito (IS) and cepstral distance (CD) that can quantify the degradation in the speech caused due to the reverberation. Table-\ref{tab:quality} shows the simulated distant speech signals  generated using real recordings of RIRs in various acoustic environments are heavily distorted w.r.t clean speech signals from WSJ corpus. 

\begin{table}[h!]
  \begin{center}
  \begin{tabular}{|p{25mm}|p{13mm}|p{10mm}|p{9mm}|p{8mm}|p{8mm}|}
  \hline 
  \textbf{Data} & \textbf{SNR(dB)} & \textbf{PESQ} & \textbf{IS} & \textbf{CD}  \\ \hline
  Lecture Hall & -3.18 & 1.52 & 8.11 & 6.22 \\ 
  Office Room & -3.09 & 1.62 & 3.5 & 5.3 \\ 
  Meeting Room & -2.58 & 2.33 & 7.98 & 4.84 \\ 
  Stairway & -2.69 & 1.87 & 12.06 & 5.59 \\ \hline
  \end{tabular}
  \caption{Objective Quality Measures for simulated distant speech signals w.r.t. clean speech signals from WSJ corpus.\vspace{1em}}
  \vspace{-2em}
  \label{tab:quality}
  \end{center}
\end{table}

\begin{figure}[h!]
  \centering
 m  \includegraphics[height=3.8cm, width=\linewidth]{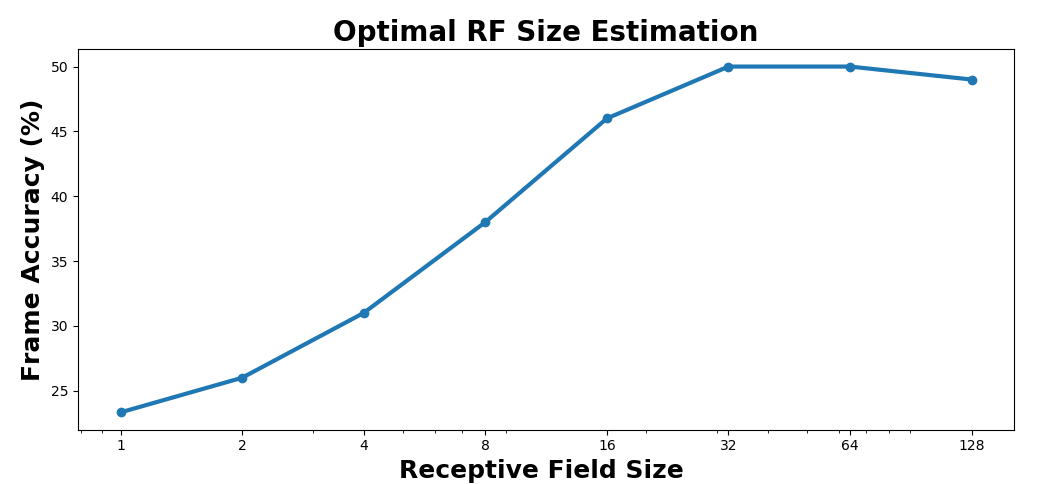}
  \caption{Optimal Kernel Size for capturing long-term dynamics in simulated distant speech}
  \label{fig:kernel}
  \vspace{-1em}
\end{figure}

Next, we run an elementary empirical experiment using a standard CNN with one layer of convolution to comprehend "how long?" is actually long enough to capture the long-term dynamics in distant simulated speech signals, We train this single layer standard CNN for various receptive field sizes\footnote{For a single layer standard CNN, kernel size will be the same as the receptive field size} using the simulated speech signals, see Fig-\ref{fig:kernel}. It is evident from this experiment that the accuracy increases with an increase in RF size. However, having a greater RF size than required neither hurts nor improves the system's performance. Thus, for our experiments, we fix the number of parameters across all the networks based on the optimal RF size determined from this experiment.
\vspace{-0.5em}

\vspace{-0.5em}

\begin{table}[h!]
  \begin{tabular}{|p{25mm}|p{30mm}|p{10mm}|}
  \hline 
  \textbf{Network} & \textbf{Architecture} & \textbf{Acc(\%)}  \\ \hline
  \multirow{4}{*}{Standard CNN}  & W:5 L:6 & 55.09  \\
  & W:5 L:8  & 57.92  \\
  & W:5 L:10 & 60.52  \\
  & W:6 L:10 & 60.47  \\
  & \textbf{W:7 L:10} & \textbf{61.80}  \\\hline
  \multirow{4}{*}{Dilated Net}  & W:5 d:2 L:5 & 56.59   \\
  & W:5 d:2 L:7 & 58.41  \\
  & \textbf{W:5 D:4 L:7} & \textbf{64.17}  \\
  & W:5 D:4 L:9 & 63.65  \\
  & W:5 D:8 L:7 & 61.42  \\ \hline
  \multirow{3}{*}{Recursive Net}  & EMBD Layer:1 & 60.23  \\
  & \textbf{EMBD Layer:2} & \textbf{61.45}  \\
   (RIN/ROUT:3)  & EMBD layer:3 & 60.55  \\ \hline
  \multirow{4}{*}{Hourglass CNN}& HG:1 W:5 L:3 & 63.21  \\
  & HG:3 W:3 L:3 & 65.98  \\
  & HG:3 W:3 L:5 & 67.25  \\
  & \textbf{HG:3 W:5 L:5} & \textbf{67.55}  \\
  & HG:5 W:3 L:5 & 67.48  \\
  & HG:5 W:5 L:5 & 67.01  \\ \hline
  \end{tabular}
  \caption{Performance of Standard CNN and large receptive field networks for different configurations.\vspace{-0.5em}}
  \label{tab:studytable}
\end{table}%

Furthermore, for better understanding of LRF networks, we test the frame accuracies obtained by all the networks on Dev93 for various architectures, see Table-\ref{tab:studytable}. It shows the performance of all LRF networks. We observe a linearly growing trend in standard CNN's efficiency (in terms of validation frame accuracy) with increased kernel size and number of layers, in other words, RF size. Unlike the standard CNNs, the LRF networks showed optimal performance over all the variations tested in their architectures for a specific RF size. It can, therefore, be expressed that having a large receptive field customized to distortion levels in speech can enhance the efficiency of a system; LRF networks can achieve this at a reduced computational expense than standard CNNs.

\begin{table}[h!]
  \centering
  \begin{tabular}{|l|c|c|c|c|}
  \hline 
  \multirow{2}{*}{\textbf{Network}} & \multicolumn{2}{c|}{\textbf{Frame Acc. (\%)}} & \multicolumn{2}{c|}{\textbf{WER(\%)}}  \\ \cline{2-5}
  & \textbf{Clean} & \textbf{Reverb} & \textbf{Clean} & \textbf{Reverb} \\ \hline
  Standard CNN  & 71.39 & 60.48 & 8.13 & 18.31 \\
  Dilated Net   & 74.16 & 64.17 & \textbf{7.25} & 17.52 \\
  Recursive Net & 73.61 & 65.17 & 7.54 & 17.13 \\
  Hourglass Net & \textbf{75.43} & \textbf{67.00} & 7.98 & \textbf{16.68} \\ \hline
  \end{tabular}
  \caption{WER and frame accuracies of LRF networks for clean and simulated distant speech versions of Eval93 (with fixed number of parameters $\approx 25600$).\vspace{-1em}}
  \label{tab:results}
\end{table}%

We observe that all LRF networks have minor relative improvements in performance compared to the standard CNNs for clean speech signals. However, for distant speech signals, where the reverberation introduces smearing effects in both time and frequency, we see higher relative improvements using the LRF networks compared to a standard CNN for a fixed number of parameters in order to reduce the architectural complexity, see Table-\ref{tab:results}. This indicates the importance of capturing the long-term dynamics for distant speech recognition. Although the dilated networks have the best WER for clean speech, it can be argued that the architectures chosen in this comparison study were forced to have the same number of learning parameters instead of being the best in their respective category. Nonetheless, the best WER performance for the distant speech was achieved by Hourglass network. 
\vspace{-0.8em}

\section{Conclusion}
This paper highlights the importance of capturing long-term temporal dependencies of the speech signal in distant speech recognition systems. We begin by understanding the importance of the receptive field and its role in convolutional neural networks. We then study and compare a conventional CNN with dilated and variants of large receptive field networks. We used clean speech signals from WSJ corpus to simulate distant speech signals with real recordings of RIRs. Later, we did analyze the impacts of reverberation on speech using quality measures such as SNR, PESQ, Itakura-Saito and cepstral distance. We also studied convolutional CNNs with various receptive field size to better understand its impact on distant speech. Using the optimal RF size, we then compare the LRF networks constraining the parameters to find that hourglass network performs 1.8\% and 8.9\% relatively better compared to standard CNNs for clean and distant speech signals. As End-to-End speech systems have gained increasing attention in recent years, future works will explore the performance of the LRF networks for End-to-End systems.
\vfill
\newpage


\bibliographystyle{IEEEbib}
\bibliography{strings,refs}

\end{document}